\documentclass{article}

\usepackage{theorem}
\usepackage{latexsym}
\usepackage{amssymb}

\def\disp{\displaystyle}
\def\aa{\alpha}
\def\dl{\delta}
\def\ep{\varepsilon}
\def\ka{\kappa}
\def\Gm{\Gamma}
\def\Ai{\mbox{Ai}}
\def\Bi{\mbox{Bi}}
\def\ra{\rightarrow}
\def\da{\downarrow}
\def\lora{\longrightarrow}

\newcommand{\BZ}{\mathbb Z}

\newcommand{\ms}[1]{\mbox{\small{#1}}}
\newcommand{\BTqp}{
\begin{table}
\caption{\label{tbl-BT6}B\"acklund transformations of P$_{\rm VI}$}
\center{
\renewcommand{\arraystretch}{1.6}
\arraycolsep=2.0pt
$\begin{array}{|c||ccccc|cc|}
\hline
& \aa_0 & \aa_1 & \aa_2 & \aa_3 & \aa_4 
& f_4 & f_2\\
\hline\hline
\ s_0\ &-\aa_0& \aa_1&\aa_2+\aa_0& \aa_3& \aa_4
&f_4&\disp f_2-\frac{\aa_0}{f_0} \\
s_1& \aa_0&-\aa_1&\aa_2+\aa_1& \aa_3& \aa_4&f_4&f_2 \\
s_2&\ \aa_0+\aa_2 &\aa_1+\aa_2&-\aa_2& \aa_3+\aa_2&\aa_4+\aa_2\
&\disp \ f_4+\frac{\aa_2}{f_2}&f_2 \\
s_3& \aa_0& \aa_1&\aa_2+\aa_3&-\aa_3& \aa_4
&f_4&\disp f_2-\frac{\aa_3}{f_3} \\
s_4& \aa_0& \aa_1&\aa_2+\aa_4& \aa_3&-\aa_4
&f_4&\disp f_2-\frac{\aa_4}{f_4} \\
\hline
s_5&\aa_1&\aa_0&\aa_2&\aa_4&\aa_3
&\disp t\frac{f_3}{f_0}&\disp -\frac{f_0(f_2f_0+\aa_2)}{t(t-1)}\\
s_6&\aa_3&\aa_4&\aa_2&\aa_0&\aa_1
&\disp \frac{t}{f_4}&\disp -\frac{f_4(f_4f_2+\aa_2)}{t}\\
s_7&\aa_4&\aa_3&\aa_2&\aa_1&\aa_0
&\disp \frac{f_0}{f_3}&\disp \frac{f_3(f_3f_2+\aa_2)}{t-1}\\
\hline
\end{array}$
}
\end{table}
}

\theoremstyle{plain} \newtheorem{thm}{Theorem}[section]
\theoremstyle{plain} \newtheorem{prop}[thm]{Proposition}
\theoremstyle{plain} 
\theoremstyle{plain} \newtheorem{rem}[thm]{Remark}
\theoremstyle{plain} \newtheorem{coro}[thm]{Corollary}
\theoremstyle{plain} 

\makeatletter
 
 \@addtoreset{equation}{section}
\makeatother

\begin{document}

\title{Classical transcendental solutions of the Painlev\'e equations
and their degeneration}
\author{Tetsu Masuda \\
Department of Mathematics, Kobe University, \\
Rokko, Kobe, 657-8501, Japan \\
{\small masuda@math.kobe-u.ac.jp}}
\date{}

\maketitle

\begin{abstract}
We present a determinant expression for a family of classical 
transcendental solutions of the Painlev\'e V and the Painlev\'e VI 
equation. 
Degeneration of these solutions along the process of coalescence for the 
Painlev\'e equations is discussed. 
\end{abstract}

\section{Introduction}
It is known that each of the Painlev\'e equations (except for 
P$_{\rm I}$) admit two classes of classical solutions. 
One is classical transcendental solutions expressible in terms of
special functions of hypergeometric type. 
Another one is algebraic or rational solutions. 
It is also known that such classical solutions are located on special 
places from a viewpoint of symmetry of the affine Weyl group as a
B\"acklund transformations group of each Painlev\'e equation. 
The rough picture is that classical transcendental solutions exist on 
the reflection hyperplanes of the affine Weyl group and algebraic (or 
rational) solutions do on the fixed points with respect to the
B\"acklund transformations corresponding to automorphisms of the Dynkin
diagram. 

In this paper, we concentrate our attention on the classical
transcendental solutions. 
One of the important features of these solutions is that they can be
expressed in terms of 2-directional Wronskians or Casorati determinants 
whose entries are given by the corresponding special function of
hypergeometric type. 
In fact, such determinant expressions for P$_{\rm II}$, P$_{\rm III}$
and P$_{\rm IV}$ have been presented~\cite{O3,O4,P4:rat,NY:P4}. 

The aim of this paper is to present a determinant expression for a
family of classical transcendental solutions of the Painlev\'e V
equation 
\begin{equation}
\frac{d^2q}{dt^2}=
\left(\frac{1}{2q}+\frac{1}{q-1}\right)\left(\frac{dq}{dt}\right)^2
-\frac{1}{t}\frac{dq}{dt}
+\frac{(q-1)^2}{2t^2}\left(\ka_{\infty}^2 q-\frac{\ka_0^2}{q}\right)
-(\theta+1)\frac{q}{t}-\frac{q(q+1)}{2(q-1)}, 
\end{equation}
which is equivalent to the Hamiltonian system 
\begin{equation}
\hskip-40pt
\mbox{S$_{\rm V}$~:} \hskip30pt
q'=\frac{\partial H}{\partial p}, \quad p'=-\frac{\partial H}{\partial q}, \quad
'=t\frac{d}{dt}, 
\end{equation}
with the Hamiltonian 
\begin{equation}
\begin{array}{c}
\medskip
\disp
H=q(q-1)^2p^2-\left[\ka_0(q-1)^2+\theta q(q-1)+tq\right]p+\ka(q-1), \\
\ka=\frac{1}{4}(\ka_0+\theta)^2-\frac{1}{4}\ka_{\infty}^2, 
\end{array}
\end{equation}
and the Painlev\'e VI equation 
\begin{equation}
\begin{array}{l}
\disp
 \frac{d^2q}{dt^2}=
 \frac{1}{2}\left(\frac{1}{q}+\frac{1}{q-1}+\frac{1}{q-t}\right)\left(\frac{dq}{dt}\right)^2
 -\left(\frac{1}{t}+\frac{1}{t-1}+\frac{1}{q-t}\right)\frac{dq}{dt} \\ \\
\disp
 \hskip50pt+\frac{q(q-1)(q-t)}{2t^2(t-1)^2}
 \left[\ka_{\infty}^2-\ka_0^2\frac{t}{q^2}
      +\ka_1^2\frac{t-1}{(q-1)^2}+(1-\theta^2)\frac{t(t-1)}{(q-t)^2}\right],
\end{array}
\end{equation}
which is equivalent to the Hamiltonian system 
\begin{equation}
\hskip-40pt
\mbox{S$_{\rm VI}$~:} \hskip30pt
q'=\frac{\partial H}{\partial p}, \quad p'=-\frac{\partial H}{\partial q}, \quad
'=t(t-1)\frac{d}{dt}, \label{cano6}
\end{equation}
with the Hamiltonian 
\begin{equation}
\begin{array}{c}
\medskip
\disp
H=q(q-1)(q-t)p^2-[\ka_0(q-1)(q-t)+\ka_1q(q-t)+(\theta-1)q(q-1)]p+\ka(q-t), \\
\ka=\frac{1}{4}(\ka_0+\ka_1+\theta-1)^2-\frac{1}{4}\ka_{\infty}^2, 
\end{array}   \label{H6}
\end{equation}
respectively. 

Let us explain how to construct a family of classical transcendental
solutions of the Painlev\'e equations. 
As an example, we take P$_{\rm II}$ : 
\begin{equation}
\frac{d^2q}{dt^2}=2q^3-2tq+2\left(\aa+\frac{1}{2}\right), 
\end{equation}
which is equivalent to the Hamiltonian system 
\begin{equation}
\hskip-40pt
\mbox{S$_{\rm II}$~:} \hskip30pt
q'=\frac{\partial H}{\partial p}, \quad p'=-\frac{\partial H}{\partial q}, \quad
'=\frac{d}{dt}, \label{cano2}
\end{equation}
with the Hamiltonian 
\begin{equation}
H=-p^2-(q^2-t)p+\aa q. \label{H2}
\end{equation}
If $\aa=0$, the right-hand side of the second equation of (\ref{cano2})
is divisible by $p$. 
This means that S$_{\rm II}$ admits the specialization of $p=0$ if
$\aa=0$. 
The first equation of (\ref{cano2}) yields the Riccati equation
$q'=-q^2+t$. 
Setting $q=(\log\varphi)'$, we get a linear equation
$\varphi''=t\varphi$, which is nothing but Airy's differential
equation. 
Thus, we find that P$_{\rm II}$ admits, when $\aa=0$, the 1-parameter
family of special solutions expressed by a rational function of the Airy
function and its derivative. 

Let introduce the $\tau$-function via the Hamiltonian (\ref{H2}) as
$H(\aa)=(\log\tau(\aa))'$. 
Then, it is known that a sequence of the $\tau$-functions
$\tau_n=\tau(\aa+n)~(n\in\BZ)$ satisfies the Toda equation 
\begin{equation}
\tau_{n+1}\tau_{n-1}=\tau_n''\tau_n-(\tau_n')^2,  \label{Toda:P2}
\end{equation}
which corresponds to the B\"acklund transformation 
\begin{equation}
\aa\to\aa-1, \quad q\to -q-\frac{2\aa}{q'+q^2-t}. 
\end{equation}
We iterate the B\"acklund transformations to the above Riccati solution
to obtain a family of classical transcendental solutions. 
What we have to do is reduced to solving the Toda equation
(\ref{Toda:P2}) with the initial conditions 
\begin{equation}
\tau_{-1}=0, \quad \tau_0=1, \quad \tau_1=\varphi. 
\end{equation}
Therefore, we have the following proposition~\cite{O3}. 

\begin{prop}\label{spec:P2}(Okamoto)~
We define the functions $\tau_n~(n\in\BZ_{\ge 0})$ by 
\begin{equation}
\tau_n =
 \left|
  \begin{array}{cccc}
   \varphi^{(0)}   & \varphi^{(1)} & \cdots & \varphi^{(n-1)}  \\
   \varphi^{(1)}   & \varphi^{(2)} & \cdots & \varphi^{(n)}    \\
   \vdots          & \vdots        & \ddots & \vdots           \\
   \varphi^{(n-1)} & \varphi^{(n)} & \cdots & \varphi^{(2n-2)}
  \end{array}
 \right|,    \quad \varphi^{(k)}=\left(\frac{d}{dt}\right)^k\varphi, \label{tau:P2}
\end{equation}
where $\varphi$ is the general solution of Airy's differential equation 
$\varphi''=t\varphi$. 
Then, 
\begin{equation}
q=\frac{d}{dt}\log\frac{\tau_{n+1}}{\tau_n}, \quad 
p=\frac{\tau_{n+1}\tau_{n-1}}{\tau_n^2}, 
\end{equation}
\begin{equation}
\aa=n,
\end{equation}
gives a family of classical transcendental solutions of S$_{\rm II}$. 
\end{prop}

By the similar procedures, it is possible to obtain a determinant 
expression for a family of classical transcendental solutions to 
P$_{\rm V}$ and P$_{\rm VI}$, which are presented in Section \ref{PV}
and \ref{PVI}, respectively. 
As is well known, P$_{\rm VI}$ degenerates P$_{\rm V},\ldots$, 
P$_{\rm I}$ by successive limiting procedures~\cite{Pa,GtoP}. 
In section \ref{deg}, we discuss the degeneration of the family of
classical transcendental solutions.

\section{Classical transcendental solutions of the Painlev\'e V equation\label{PV}}
\subsection{The symmetric form of the Painlev\'e V equation}
First, we summarize the symmetric form of the Painlev\'e V
equation~\cite{NY1,NY2}. 
The symmetric form of P$_{\rm V}$ is given by 
\begin{equation}
\begin{array}{l}
\medskip
\disp 
 f_0'=f_0 f_2(f_1-f_3)+\left(\frac{1}{2}-\aa_2\right)f_0+\aa_0 f_2, \\
\medskip
\disp 
 f_1'=f_1 f_3(f_2-f_0)+\left(\frac{1}{2}-\aa_3\right)f_1+\aa_1 f_3, \\
\medskip
\disp 
 f_2'=f_2 f_0(f_3-f_1)+\left(\frac{1}{2}-\aa_0\right)f_2+\aa_2 f_0, \\
\disp 
 f_3'=f_3 f_1(f_0-f_2)+\left(\frac{1}{2}-\aa_1\right)f_3+\aa_3 f_1,
\end{array}   \quad '=t\frac{d}{dt}, \label{sym:A3}
\end{equation}
with the normalization conditions 
\begin{equation}
\aa_0+\aa_1+\aa_2+\aa_3=1, \quad f_0+f_2=f_1+f_3=\sqrt{t}. 
\end{equation}
The correspondence to the canonical variables and parameters of 
S$_{\rm V}$ is given by 
\begin{equation}
q=-\frac{f_3}{f_1}, \quad p=\frac{1}{\sqrt{t}}f_1(f_0f_1+\aa_0), 
\end{equation}
and 
\begin{equation}
\ka_{\infty}=\aa_1, \quad \ka_0=\aa_3, \quad \theta=\aa_2-\aa_0-1, 
\end{equation}
respectively. 
The B\"acklund transformations of P$_{\rm V}$ are described as follows, 
\begin{equation}
\begin{array}{lll}
\medskip
{\disp s_i(\aa_i)=-\aa_i,} & {\disp s_i(\aa_j)=\aa_j+\aa_i~~(j=i \pm 1),} & 
{\disp s_i(\aa_j)=\aa_j~~(j \ne i,i \pm 1),} \\ 
\medskip
{\disp s_i(f_i)=f_i,} & {\disp s_i(f_j)=f_j \pm \frac{\aa_i}{f_i}~~(j=i \pm 1),} &
{\disp s_i(f_j)=f_j ~~(j \ne i,i \pm 1),} \\
{\disp \pi(\aa_j)=\aa_{j+1},} & \pi(f_j)=f_{j+1}, & 
\end{array}
\end{equation}
where the subscripts $i=0,1,2,3$ are understood as elements of
$\BZ/4\BZ$. 
The Hamiltonians $h_i$ of the system (\ref{sym:A3}) are given by 
\begin{equation}
\begin{array}{l}
\medskip
\disp
 h_0=f_0 f_1 f_2 f_3
     +\frac{\aa_1+2\aa_2-\aa_3}{4}f_0 f_1+\frac{\aa_1+2\aa_2+3\aa_3}{4}f_1 f_2 \\
\disp \hskip60pt
     -\frac{3\aa_1+2\aa_2+\aa_3}{4}f_2 f_3+\frac{ \aa_1-2\aa_2-\aa_3}{4}f_3 f_0
     +\frac{(\aa_1+\aa_3)^2}{4},
\end{array}
\end{equation}
and $h_i=\pi^i(h_0)$. 
Then, we have 
\begin{equation}
s_i(h_j)=h_j~(i \ne j), \quad s_i(h_i)=h_i+\sqrt{t}~\frac{\aa_i}{f_i}, 
\quad \pi(h_i)=h_{i+1}. \label{BT:h:P5}
\end{equation}
Introducing the $\tau$-functions $\tau_i$ as 
\begin{equation}
h_i=(\log\tau_i)', 
\end{equation}
we find that the B\"acklund transformations for the $\tau$-functions are
described as 
\begin{equation}
s_i(\tau_j)=\tau_j~~(i\neq j), \quad s_i(\tau_i)=f_i\frac{\tau_{i-1}\tau_{i+1}}{\tau_i}, 
\quad \pi(\tau_i)=\tau_{i+1}. \label{BT:tau:P5}
\end{equation}
The canonical variables of S$_{\rm V}$ are recovered from the $\tau$-functions by 
\begin{equation}
q=-\frac{\tau_3 s_3(\tau_3)}{\tau_1 s_1(\tau_1)}, \quad 
p=\frac{1}{\sqrt{t}}\frac{\tau_1s_1(\tau_1)s_0s_1(\tau_1)}{\tau_2^2\tau_3}. 
\label{qp:tau:P5}
\end{equation}

Let us define the translation operators $T_i$ by 
\begin{equation}
T_1=\pi s_3s_2s_1, \quad T_2=s_1\pi s_3s_2, \quad 
T_3=s_2s_1\pi s_3, \quad T_0=s_3s_2s_1\pi,
\end{equation}
which commute with each other and act on parameters $\aa_i$ as 
\begin{equation}
T_i(\aa_{i-1})=\aa_{i-1}+1, \quad T_i(\aa_i)=\aa_i-1, \quad 
T_i(\aa_j)=\aa_j~(j \ne i-1,i).
\end{equation}
Noting that $T_1T_2T_3T_0=1$, we set 
\begin{equation}
\tau_{k,l,m}=T_1^kT_2^lT_3^m(\tau_0), \quad k,l,m \in \BZ. 
\end{equation}
Then, from (\ref{BT:tau:P5}) and (\ref{qp:tau:P5}), we have 
\begin{equation}
\begin{array}{l}
\medskip
\disp 
T_1^kT_2^lT_3^m(f_0)=
\frac{\tau_{k,l,m}\tau_{k+2,l+1,m+1}}
     {\tau_{k+1,l+1,m+1}\tau_{k+1,l,m}}, \quad 
T_1^kT_2^lT_3^m(f_1)=
\frac{\tau_{k+1,l,m}\tau_{k,l+1,m}}
     {\tau_{k,l,m}\tau_{k+1,l+1,m}}, \\
\disp 
T_1^kT_2^lT_3^m(f_2)=
\frac{\tau_{k+1,l+1,m}\tau_{k+1,l,m+1}}
     {\tau_{k+1,l,m}\tau_{k+1,l+1,m+1}}, \quad 
T_1^kT_2^lT_3^m(f_3)=
\frac{\tau_{k+1,l+1,m+1}\tau_{k,l,m-1}}
     {\tau_{k+1,l+1,m}\tau_{k,l,m}}, 
\end{array}
\end{equation}
and 
\begin{equation}
T_1^kT_2^lT_3^m(q)=
-\frac{\tau_{k+1,l+1,m+1}\tau_{k,l,m-1}}
      {\tau_{k+1,l,m}\tau_{k,l+1,m}}, \quad 
T_1^kT_2^lT_3^m(p)=
\frac{1}{\sqrt{t}}
\frac{\tau_{k+1,l,m}\tau_{k,l+1,m}\tau_{k+2,l+2,m+1}}
     {\tau_{k+1,l+1,m}^2\tau_{k+1,l+1,m+1}}, 
\end{equation}
respectively. 
It is possible to derive the Toda equation with respect to each
translation operator. 
For the $T_1$-direction, we have 
\begin{equation}
\tau_{k+1,l,m}\tau_{k-1,l,m}=\frac{1}{\sqrt{t}}
\left[(\log\tau_{k,l,m})''+\frac{3\aa_1+2\aa_2+\aa_3-3k+l+m}{4}t\right]
\tau_{k,l,m} \cdot \tau_{k,l,m}. \label{Toda:P5}
\end{equation}

\subsection{A Riccati solution}
If $\aa_0=0$, the right-hand side of the first equation of
(\ref{sym:A3}) is divisible by $f_0$. 
This means that the system (\ref{sym:A3}) admits the specialization of
$f_0=0$ if $\aa_0=0$. 
Then, setting $f_1=\sqrt{t}f$, we see that $f$ satisfies a Riccati equation 
\begin{equation}
f'=tf(1-f)-(\aa_1+\aa_3)f+\aa_1. 
\end{equation}
By the dependent variable transformation 
$\disp f=\frac{d}{dt}\log\varphi$, 
we have for $\varphi$ the linear equation 
\begin{equation}
\left[t\frac{d^2}{dt^2}+(\aa_1+\aa_3-t)\frac{d}{dt}-\aa_1\right]\varphi=0, \label{CHG}
\end{equation}
which is nothing but the confluent hypergeometric differential equation.
We set $\aa_1=a$ and $\aa_3=c-a$, and impose ``non-integer conditions''
on them appropriately. 
The general solution of (\ref{CHG}) is expressed by 
\begin{equation}
\varphi=c_1\frac{\Gm(a)\Gm(c-a)}{\Gm(c)}F(a,c;t)
       +c_2\frac{1}{\sin\pi(c-a)\Gm(2-c)}t^{1-c}F(a-c+1,2-c;t), 
\end{equation}
where $F(a,c;t)$ denotes Kummer's confluent hypergeometric function, and
$c_i~(i=1,2)$ are arbitrary complex constants. 
For simplicity, we denote 
\begin{equation}
f_{i,j}=F\ms{$(a+i,c+j;t)$}, \quad g_{i,j}=t^{1-c-j}F\ms{$(a-c+1+i-j,2-c-j;t)$}
\quad (i,j\in\BZ). 
\end{equation}
By using the contiguity relations of Kummer's function, 
we obtain the following proposition. 

\begin{prop}\label{P5:Ric}
We define the functions $\varphi_{i,j}$ by 
\begin{equation}
\varphi_{i,j}=c_1\frac{\Gm(a+i)\Gm(c-a-i+j)}{\Gm(c+j)}f_{i,j}
       +c_2\frac{1}{\sin\pi(c-a-i+j)\Gm(2-c-j)}g_{i,j}.      \label{entry:5}
\end{equation}
Then, 
\begin{equation}
(f_0,f_1,f_2,f_3)= 
\left(0,\sqrt{t}~\frac{\varphi_{1,1}}{\varphi_{0,0}},
      \sqrt{t},\sqrt{t}\frac{\varphi_{0,1}}{\varphi_{0,0}}\right), \quad 
(\aa_0,\aa_1,\aa_2,\aa_3)=(0,a,1-c,c-a),  \label{Ric:N5}
\end{equation}
and 
\begin{equation}
q=-\frac{\varphi_{0,1}}{\varphi_{1,1}}, \quad p=0, \qquad 
\ka_{\infty}=a, \quad \ka_0=c-a, \quad \theta=-c, 
\end{equation}
give a Riccati solution of the symmetric form of P$_{\rm V}$ and the
Hamiltonian system S$_{\rm V}$, respectively. 
\end{prop}

\subsection{A Determinant formula for a family of classical transcendental solutions}
First, we calculate the Hamiltonians and $\tau$-functions for the
Riccati solution in Proposition \ref{P5:Ric}. 
Under the specialization (\ref{Ric:N5}), the Hamiltonians and
$\tau$-functions are calculated as 
\begin{equation}
\begin{array}{l}
\medskip
\disp 
 h_0=\frac{\varphi_{0,0}'}{\varphi_{0,0}}-\frac{2a-c+2}{4}t+\frac{c^2}{4}, \quad 
 h_1=-\frac{2a-c-1}{4}t+\frac{(c-1)^2}{4}, \\
\disp 
 h_2=-\frac{2a-c}{4}t  +\frac{c^2}{4},     \quad 
 h_3=-\frac{2a-c+1}{4}t+\frac{(c-1)^2}{4}, 
\end{array}
\end{equation}
and 
\begin{equation}
\begin{array}{l}
\medskip
\disp 
 \tau_0=\tau_{0,0,0}=\varphi_{0,0}~t^{c^2/4}\exp\left(-\frac{2a-c+2}{4}t\right), \quad 
 \tau_1=\tau_{1,0,0}=t^{(c-1)^2/4}\exp\left(-\frac{2a-c-1}{4}t\right), \\
\medskip
\disp 
 \tau_2=\tau_{1,1,0}=t^{c^2/4}\exp\left(-\frac{2a-c}{4}t\right), \quad 
 \tau_3=\tau_{1,1,1}=t^{(c-1)^2/4}\exp\left(-\frac{2a-c+1}{4}t\right), \\
\medskip
\disp 
 s_0(\tau_0)=\tau_{2,1,1}=0, \quad 
 s_1(\tau_1)=\tau_{0,1,0}=\varphi_{1,1}t^{(c+1)^2/4}\exp\left(-\frac{2a-c+3}{4}t\right), \\
\medskip
\disp 
 s_2(\tau_2)=\tau_{1,0,1}=t^{(c-2)^2/4}\exp\left(-\frac{2a-c}{4}t\right), \\ 
\disp 
 s_3(\tau_3)=\tau_{0,0,-1}=\varphi_{0,1}t^{(c+1)^2/4}\exp\left(-\frac{2a-c+1}{4}t\right), 
\end{array}
\end{equation}
up to multiplication by some constants, respectively. 
For small $k,l,m$, we observe that $\tau_{k,l,m}$ are expressed in the
form 
\begin{equation}
\tau_{k,l,m}=\sigma_{k,l,m}t^{(c-k+l-m)^2/4-k(k-1)/2}
               \exp \left(-\frac{2a-c+2-3k+l+m}{4}t \right), \label{tau-sig:P5}
\end{equation}
with 
\begin{equation}
\sigma_{2,l,m}=0, \quad 
\sigma_{1,l,m}=\mbox{const.}, \quad 
\sigma_{0,l,m}=(\mbox{const.})\times\varphi_{l,l-m}. 
\end{equation}
Assume that $\tau_{k,l,m}$ are expressed as (\ref{tau-sig:P5}) for
any $k,l,m\in\BZ$. 
Then, the Toda equation (\ref{Toda:P5}) yields 
\begin{equation}
\sigma_{k+1,l,m}\sigma_{k-1,l,m}=
\sigma_{k,l,m}''\sigma_{k,l,m}-(\sigma_{k,l,m}')^2. 
\end{equation}
Moreover, we set 
\begin{equation}
\sigma_{k,l,m}=\omega_{k,l,m}\rho_{k,l,m}, \quad \omega_{k,l,m}=\omega_{k,l,m}(a,c), 
\end{equation}
with $\rho_{1,l,m}=1$ and $\rho_{0,l,m}=\varphi_{l,l-m}$, 
and impose that the constants $\omega_{k,l,m}$ satisfy 
\begin{equation}
\omega_{k+1,l,m}\omega_{k-1,l,m}=\omega_{k,l,m}^2. \label{omega:k:P5}
\end{equation}
Then, the functions $\rho_{k,l,m}$ are determined by the recurrence
relation 
\begin{equation}
\rho_{k+1,l,m}\rho_{k-1,l,m}=
\rho_{k,l,m}''\rho_{k,l,m}-(\rho_{k,l,m}')^2,
\end{equation}
with the initial conditions 
\begin{equation}
\rho_{2,l,m}=0, \quad \rho_{1,l,m}=1, \quad \rho_{0,l,m}=\varphi_{l,l-m}. 
\end{equation}
By Darboux's formula, the functions $\rho_{1-n,l,m}$ for
$n\in\BZ_{\ge0}$ are expressed as 
\begin{equation}
\rho_{1-n,l,m}=
 \left|
 \begin{array}{cccc}
   \varphi_{l,l-m}^{(0)}   & \varphi_{l,l-m}^{(1)} & \cdots & \varphi_{l,l-m}^{(n-1)}  \\
   \varphi_{l,l-m}^{(1)}   & \varphi_{l,l-m}^{(2)} & \cdots & \varphi_{l,l-m}^{(n)}    \\
   \vdots                  & \vdots                & \ddots & \vdots                   \\
   \varphi_{l,l-m}^{(n-1)} & \varphi_{l,l-m}^{(n)} & \cdots & \varphi_{l,l-m}^{(2n-2)}
 \end{array}
 \right|,   \quad \varphi_{l,l-m}^{(i)}=\left(t\frac{d}{dt}\right)^i \varphi_{l,l-m}. 
\end{equation}
Note that the constants $\omega_{k,l,m}$ are determined by the
recurrence relations (\ref{omega:k:P5}) and 
\begin{equation}
\begin{array}{l}
\smallskip
\omega_{1,l+1,m}\omega_{1,l-1,m}=-(a+l-1)\omega_{1,l,m}^2, \quad 
\omega_{1,l,m+1}\omega_{1,l,m-1}=(c-a-m)\omega_{1,l,m}^2, \\
\omega_{0,l+1,m}\omega_{0,l-1,m}=-(a+l-1)\omega_{0,l,m}^2, \quad 
\omega_{0,l,m+1}\omega_{0,l,m-1}=(c-a-1-m)\omega_{0,l,m}^2, 
\end{array}
\end{equation}
with the initial conditions 
\begin{equation}
\begin{array}{l}
\smallskip
\omega_{1,0,0}=\omega_{1,1,0}=\omega_{1,0,1}=\omega_{1,1,1}=1, \\
\disp
\omega_{0,0,0}=\omega_{0,1,0}=\omega_{0,0,-1}=\omega_{0,1,-1}=1. 
\end{array}
\end{equation}
Since it is possible to set $l=m=0$ without loss of generality, 
we obtain the following theorem. 

\begin{thm}\label{P5:spec}
We define the functions $\tau_n^{i,j}$ by 
\begin{equation}
\tau_n^{i,j}=
 \left|
 \begin{array}{cccc}
  \varphi_{i,j}^{(0)}   & \varphi_{i,j}^{(1)} & \cdots & \varphi_{i,j}^{(n-1)}  \\
  \varphi_{i,j}^{(1)}   & \varphi_{i,j}^{(2)} & \cdots & \varphi_{i,j}^{(n)}    \\
  \vdots            & \vdots          & \ddots & \vdots             \\
  \varphi_{i,j}^{(n-1)} & \varphi_{i,j}^{(n)} & \cdots & \varphi_{i,j}^{(2n-2)}
 \end{array}
 \right|,   \quad \varphi_{i,j}^{(k)}=\left(t\frac{d}{dt}\right)^k \varphi_{i,j}, 
\label{tau:P5}
\end{equation}
where $\varphi_{i,j}$ are given by (\ref{entry:5}). 
Then, 
\begin{equation}
\begin{array}{l}
\medskip
\disp 
f_0=\frac{1}{\sqrt{t}}
    \frac{\tau_{n+1}^{0,0}\tau_{n-1}^{1,0}}
         {\tau_n^{1,0}\tau_n^{0,0}}, \quad 
f_1=\sqrt{t}
    \frac{\tau_n^{0,0}\tau_{n+1}^{1,1}}
         {\tau_{n+1}^{0,0}\tau_n^{1,1}}, \\
\disp 
f_2=\sqrt{t}
    \frac{\tau_n^{1,1}\tau_n^{0,-1}}
         {\tau_n^{0,0}\tau_n^{1,0}}, \quad 
f_3=\sqrt{t}\left(\frac{c-a-1}{c-a}\right)^n
    \frac{\tau_n^{1,0}\tau_{n+1}^{0,1}}
         {\tau_n^{1,1}\tau_{n+1}^{0,0}}, 
\end{array}
\end{equation}
\begin{equation}
(\aa_0,\aa_1,\aa_2,\aa_3)=(-n,a+n,1-c,c-a), 
\end{equation}
and 
\begin{equation}
q=-\left(\frac{c-a-1}{c-a}\right)^n
   \frac{\tau_n^{1,0}\tau_{n+1}^{0,1}}
        {\tau_n^{0,0}\tau_{n+1}^{1,1}}, \quad 
p=-\frac{a}{c-a-1}\frac{1}{t}
   \frac{\tau_n^{0,0}\tau_{n+1}^{1,1}\tau_{n-1}^{2,1}}
        {\tau_n^{1,1}\tau_n^{1,1}\tau_n^{1,0}}, 
\end{equation}
\begin{equation}
\ka_{\infty}=a+n, \quad \ka_0=c-a, \quad \theta=-c+n, \label{para:P5}
\end{equation}
give a family of classical transcendental solutions of the symmetric 
form of P$_{\rm V}$ and the Hamiltonian system S$_{\rm V}$, 
respectively. 
\end{thm}

\begin{rem}\label{rem:P5}
Noting that we have from (\ref{BT:h:P5}) 
\begin{equation}
\sqrt{t}~\frac{\aa_i}{f_i}=t\frac{d}{dt}\log\frac{s_i(\tau_i)}{\tau_i}, 
\end{equation}
which lead us to another expression of the solutions in Theorem 
\ref{P5:spec}. 
For example, we obtain 
\begin{equation}
(a+n)\frac{\sqrt{t}}{f_1}=t\frac{d}{dt}\log\frac{\tau_{n+1}^{1,1}}{\tau_n^{0,0}}+c-t. 
\end{equation}
\end{rem}

\begin{rem}
The symmetric form of P$_{\rm V}$ admits the following symmetry 
$\sigma$: 
\begin{equation}
\begin{array}{l}
\medskip
\disp 
 \sigma(t)=-t, \\
\medskip
\disp 
 \sigma(f_0)=\sqrt{-1}f_2, \quad \sigma(f_2)=\sqrt{-1}f_0, \quad 
 \sigma(f_1)=\sqrt{-1}f_1, \quad \sigma(f_3)=\sqrt{-1}f_3, \\
\disp 
 \sigma(\aa_0)=\aa_2, \quad \sigma(\aa_2)=\aa_0, \quad 
 \sigma(\aa_1)=\aa_1, \quad \sigma(\aa_3)=\aa_3. 
\end{array}
\end{equation}
The application of $\sigma$ to the family of solutions in Theorem \ref{P5:spec} gives
another family of solutions expressed in terms of $F(a,c;-t)$. 
\end{rem}

\section{Classical transcendental solutions of the Painlev\'e VI equation\label{PVI}}
\subsection{The symmetric form of the Painlev\'e VI equation}
Here, we give a brief review of the symmetric form of the Painlev\'e VI
equation~\cite{NY3,KMNOY}. 
We set 
\begin{equation}
f_0=q-t, \quad f_3=q-1, \quad f_4=q, \quad f_2=p,
\end{equation}
and 
\begin{equation}
\aa_0=\theta, \quad \aa_1=\ka_{\infty}, \quad \aa_3=\ka_1, \quad \aa_4=\ka_0. 
\end{equation}
Then, the Hamiltonian (\ref{H6}) is written as 
\begin{equation}
H=f_2^2f_0f_3f_4-[(\aa_0-1)f_3f_4+\aa_3 f_0f_4+\aa_4 f_0f_3]f_2+\aa_2(\aa_1+\aa_2)f_0,  
\end{equation}
with 
\begin{equation}
\aa_0+\aa_1+2\aa_2+\aa_3+\aa_4=1, 
\end{equation}
and the Hamilton equation (\ref{cano6}) is written as 
\begin{equation}
\begin{array}{l}
\medskip
\disp
f_4'=2f_2f_0f_3f_4-(\aa_0-1)f_3f_4-\aa_3 f_0f_4-\aa_4 f_0f_3, \\
\medskip
\disp
f_2'=-(f_0f_3+f_0f_4+f_3f_4)f_2^2 \\
\disp
\hskip40pt
     +[(\aa_0-1)(f_3+f_4)+\aa_3(f_0+f_4)+\aa_4(f_0+f_3)]f_2-\aa_2(\aa_1+\aa_2). 
\end{array}   \label{cano6'}
\end{equation}
The fundamental B\"acklund transformations of P$_{\rm VI}$ are given in
Table \ref{tbl-BT6}. 
We define the Hamiltonians $h_i~(i=0,1,2,3,4)$ as 
\begin{equation}
h_0=H_0+\frac{t}{4},\quad h_1=s_5(H_0)-\frac{t-1}{4},\quad 
h_3=s_6(H_0)+\frac{1}{4},\quad h_4=s_7(H_0), \quad h_2=h_1+s_1(h_1), 
\end{equation}
where an auxiliary Hamiltonian $H_0$ is given by 
\begin{equation}
H_0=H+\frac{t}{4}[1+4\aa_1\aa_2+4\aa_2^2-(\aa_3+\aa_4)^2]+\frac{1}{4}[(\aa_1+\aa_4)^2+(\aa_3+\aa_4)^2+4\aa_2\aa_4]. 
\end{equation}
Introducing the $\tau$-functions $\tau_i$ as 
\begin{equation}
h_i=(\log\tau_i)', 
\end{equation}
we find that the B\"acklund transformations for the $\tau$-functions are
described as follows, 
\begin{equation}
\begin{array}{c}
\medskip
s_i(\tau_j)=\tau_j \quad (i\ne j,~~i,j=0,1,2,3,4), \\
\disp
s_i(\tau_i)=f_i\frac{\tau_2}{\tau_i} \quad (i=0,3,4), \quad 
s_1(\tau_1)=\frac{\tau_2}{\tau_1}, \quad 
s_2(\tau_2)=\frac{f_2}{\sqrt{t}}\frac{\tau_0\tau_1\tau_3\tau_4}{\tau_2}, 
\end{array}
\end{equation}
\begin{equation}
\begin{array}{lll}
\medskip
{s_5:}&{\tau_0 \ra [t(t-1)]^{\frac{1}{4}}\tau_1,}&{\tau_1 \ra [t(t-1)]^{-\frac{1}{4}}\tau_0,}\\
\medskip
      &{\tau_3 \ra t^{-\frac{1}{4}}(t-1)^{\frac{1}{4}}\tau_4,}&{\tau_4 \ra t^{\frac{1}{4}}(t-1)^{-\frac{1}{4}}\tau_3,} \\
      &{\tau_2 \ra [t(t-1)]^{-\frac{1}{2}}f_0\tau_2,}&
\end{array}
\end{equation}
\begin{equation}
s_6: \quad \tau_0 \ra i t^{\frac{1}{4}}\tau_3, \quad \tau_3 \ra -i t^{-\frac{1}{4}}\tau_0, 
     \quad \tau_1 \ra t^{-\frac{1}{4}}\tau_4,  \quad \tau_4 \ra t^{\frac{1}{4}}\tau_1, 
     \quad \tau_2 \ra t^{-\frac{1}{2}}f_4 \tau_2,
\end{equation}
\begin{equation}
\begin{array}{lll}
\medskip
{s_7:}&{\tau_0 \ra (-1)^{-\frac{3}{4}}(t-1)^{\frac{1}{4}}\tau_4,}&{\tau_4 \ra (-1)^{\frac{3}{4}}(t-1)^{-\frac{1}{4}}\tau_0,}\\
\medskip
      &{\tau_1 \ra (-1)^{\frac{3}{4}}(t-1)^{-\frac{1}{4}}\tau_3,}&{\tau_3 \ra (-1)^{-\frac{3}{4}}(t-1)^{\frac{1}{4}}\tau_1,}\\
      &{\tau_2 \ra -i(t-1)^{-\frac{1}{2}}f_3\tau_2.}&
\end{array}
\end{equation}
\BTqp

Let us define the following translation operators
\begin{equation}
\begin{array}{c}
\widehat{T}_{13}=s_1s_2s_0s_4s_2s_1s_7, \quad 
\widehat{T}_{40}=s_4s_2s_1s_3s_2s_4s_7, \\
\widehat{T}_{34}=s_3s_2s_0s_1s_2s_3s_5, \quad 
T_{14}=s_1s_4s_2s_0s_3s_2s_6,
\end{array}
\end{equation}
which act on parameters $\aa_i$ as 
\begin{equation}
\begin{array}{c}
\widehat{T}_{13}(\aa_0,\aa_1,\aa_2,\aa_3,\aa_4)=
                (\aa_0,\aa_1,\aa_2,\aa_3,\aa_4)+(0,1,0,-1,0), \\
\widehat{T}_{40}(\aa_0,\aa_1,\aa_2,\aa_3,\aa_4)=
                (\aa_0,\aa_1,\aa_2,\aa_3,\aa_4)+(-1,0,0,0,1), \\
\widehat{T}_{34}(\aa_0,\aa_1,\aa_2,\aa_3,\aa_4)=
                (\aa_0,\aa_1,\aa_2,\aa_3,\aa_4)+(0,0,0,1,-1), \\
T_{14}(\aa_0,\aa_1,\aa_2,\aa_3,\aa_4)=
      (\aa_0,\aa_1,\aa_2,\aa_3,\aa_4)+(0,1,-1,0,1). 
\end{array}
\end{equation}
Note that the action of these operators on $\tau$-functions is not
commutative. 
For example, we have 
\begin{equation}
\begin{array}{ll}
{\widehat{T}_{13}\widehat{T}_{40}(\tau_0)=-\widehat{T}_{40}\widehat{T}_{13}(\tau_0),}&
{\widehat{T}_{13}\widehat{T}_{34}(\tau_0)=-i\widehat{T}_{34}\widehat{T}_{13}(\tau_0),} \\
{\widehat{T}_{13}T_{14}(\tau_0)=iT_{14}\widehat{T}_{13}(\tau_0),}&
{\widehat{T}_{40}\widehat{T}_{34}(\tau_0)=i\widehat{T}_{34}\widehat{T}_{40}(\tau_0),} \\
{\widehat{T}_{40}T_{14}(\tau_0)=-iT_{14}\widehat{T}_{40}(\tau_0),}&
{\widehat{T}_{34}T_{14}(\tau_0)=iT_{14}\widehat{T}_{34}(\tau_0)}. 
\end{array}
\end{equation}
Setting 
\begin{equation}
\tau_{k,l,m,n}=
T_{14}^n \widehat{T}_{34}^m \widehat{T}_{40}^l \widehat{T}_{13}^k(\tau_0), \quad 
k,l,m,n \in \BZ, 
\end{equation}
we have 
\begin{equation}
\begin{array}{l}
\medskip
\disp
T_{14}^n\widehat{T}_{34}^m\widehat{T}_{40}^l\widehat{T}_{13}^k(f_0)=
-it^{\frac{1}{2}}(t-1)^{\frac{1}{2}}
\frac{\tau_{k,l,m,n}\tau_{k-1,l-2,m-1,n+1}}
     {\tau_{k-1,l-1,m-1,n}\tau_{k,l-1,m,n+1}}, \\
\medskip
\disp
T_{14}^n\widehat{T}_{34}^m\widehat{T}_{40}^l\widehat{T}_{13}^k(f_3)=
i(t-1)^{\frac{1}{2}}
\frac{\tau_{k,l-1,m-1,n}\tau_{k-1,l-1,m,n+1}}
     {\tau_{k-1,l-1,m-1,n}\tau_{k,l-1,m,n+1}}, \\
\medskip
\disp
T_{14}^n\widehat{T}_{34}^m\widehat{T}_{40}^l\widehat{T}_{13}^k(f_4)=
t^{\frac{1}{2}}
\frac{\tau_{k,l-1,m,n}\tau_{k-1,l-1,m-1,n+1}}
     {\tau_{k-1,l-1,m-1,n}\tau_{k,l-1,m,n+1}}, \\
\disp
T_{14}^n\widehat{T}_{34}^m\widehat{T}_{40}^l\widehat{T}_{13}^k(f_2)=
-(t-1)^{-\frac{1}{2}}
\frac{\tau_{k-1,l-1,m-1,n}\tau_{k,l-1,m,n+1}\tau_{k+1,l,m,n-1}}
     {\tau_{k,l,m,n}\tau_{k,l-1,m-1,n}\tau_{k,l-1,m,n}}. 
\end{array}
\end{equation}
It is possible to derive the Toda equation with respect to each
translation operator. 
For the $T_{14}$-direction, we get 
\begin{equation}
\begin{array}{l}
\medskip
\disp
\tau_{k,l,m,n+1}\tau_{k,l,m,n-1} \\
\disp \hskip10pt
=-t^{-\frac{1}{2}}
\left[(t-1)\frac{d}{dt}(\log\tau_{k,l,m,n})'-(\log\tau_{k,l,m,n})'
      +\frac{(\aa_1+\aa_4+k+l-m+2n)^2}{4}+\frac{1}{2}\right]\tau_{k,l,m,n}^2. 
\end{array}   \label{Toda:P6}
\end{equation}

\subsection{A Riccati solution}
If $\aa_2=0$, it is possible to specialize $f_2=0$. 
Then, the Hamilton equation (\ref{cano6'}) yields 
\begin{equation}
f_4'=-(\aa_0-1)f_3f_4-\aa_3f_0f_4-\aa_4f_0f_3, 
\end{equation}
which is equivalent to a Riccati equation 
\begin{equation}
q'=\aa_1 q^2+[(\aa_3+\aa_4)t+(\aa_0+\aa_4-1)]q-\aa_4 t. 
\end{equation}
We set 
\begin{equation}
\aa_0=-b, \quad \aa_1=a, \quad \aa_3=c-a, \quad \aa_4=b-c+1,
\end{equation}
and impose ``non-integer conditions'' on them appropriately. 
By the dependent variable transformation 
\begin{equation}
aq=-\frac{\varphi'}{\varphi}-(b+1)t+c,
\end{equation}
we have for $\varphi$ the linear equation 
\begin{equation}
t(t-1)\frac{d^2\varphi}{dt^2}+[(a+b+3)t-(c+1)]\frac{d\varphi}{dt}+(a+1)(b+1)\varphi=0, 
\label{HG}
\end{equation}
which is nothing but the hypergeometric differential equation. 
The general solution of (\ref{HG}) is expressed by 
\begin{equation}
\begin{array}{l}
\smallskip
\disp
\varphi=c_1\frac{\Gm(a+1)\Gm(b+1)}{\Gm(c+1)}F(a+1,b+1,c+1;t) \\
\disp \hskip30pt
       +c_2\frac{\Gm(a-c+1)\Gm(b-c+1)}{\Gm(1-c)}t^{-c}F(a-c+1,b-c+1,1-c;t), 
\end{array}
\end{equation}
where $F(a,b,c;t)$ denotes Gauss's hypergeometric function and
$c_i~(i=1,2)$ are arbitrary complex constants. 
For simplicity, we denote 
\begin{equation}
\begin{array}{l}
\medskip
\disp
f^{kl}_m=F\ms{$(a+k,b+l,c+m;t)$}, \\
\disp
g^{kl}_m=t^{1-c-m}F\ms{$(a-c+1+k-m,b-c+1+l-m,2-c-m;t)$}
\end{array} \quad (k,l,m\in\BZ). 
\end{equation}
By using the contiguity relations of Gauss's hypergeometric function, 
we obtain the following proposition. 

\begin{prop}\label{P6:Ric}
We define the function $\varphi_{k,l,m}$ by 
\begin{equation}
\begin{array}{l}
\smallskip
\disp
\varphi_{k,l,m}=c_1\frac{\Gm(a+k+1)\Gm(b+l+2)}{\Gm(c+m+1)}f^{k+1,l+2}_{m+1} \\
\disp \hskip40pt 
+c_2\frac{\Gm(a-c+1+k-m)\Gm(b-c+2+l-m)}{\Gm(1-c-m)}g^{k+1,l+2}_{m+1}. 
\end{array}   \label{entry:6}
\end{equation}
Then, 
\begin{equation}
\begin{array}{c}
\medskip
\disp
f_0=b\frac{\varphi_{-1,-2,-1}}{\varphi_{0,-1,0}}, \quad 
f_3=(c-a)\frac{\varphi_{-1,-1,0}}{\varphi_{0,-1,0}}, \quad 
f_4=\frac{\varphi_{-1,-1,-1}}{\varphi_{0,-1,0}}, \quad 
f_2=0, \\
\disp
\aa_0=-b, \quad \aa_1=a, \quad \aa_3=c-a, \quad \aa_4=b-c+1,
\end{array}   \label{Ric:N6}
\end{equation}
gives a Riccati solution of the symmetric form of P$_{\rm VI}$. 
\end{prop}

\subsection{A Determinant formula for a family of classical transcendental solutions}
First, we calculate the Hamiltonians and $\tau$-functions for the
Riccati solution in Proposition \ref{P6:Ric}. 
Under the specialization (\ref{Ric:N6}), the Hamiltonians are calculated as 
\begin{equation}
h_i=A_i(t-1)+B_it, \quad (i=0,1,3,4),
\end{equation}
with 
\begin{equation}
\begin{array}{ll}
\smallskip
{A_0=-\frac{1}{4}(a+b-c+1)^2-\frac{1}{4}(a-b-1)^2,} & 
{B_0=\frac{1}{4}(a+b-c+1)^2+\frac{1}{2},} \\
\smallskip
{A_1=-\frac{1}{4}(a+b-c)^2-\frac{1}{4}(a-b-1)^2-\frac{1}{4},} & 
{B_1=\frac{1}{4}(a+b-c)^2+\frac{1}{4},} \\
\smallskip
{A_3=-\frac{1}{4}(a+b-c+1)^2-\frac{1}{4}(a-b)^2-\frac{1}{4},} & 
{B_3=\frac{1}{4}(a+b-c+1)^2+\frac{1}{2},} \\
{A_4=-\frac{1}{4}(a+b-c)^2-\frac{1}{4}(a-b)^2,} & 
{B_4=\frac{1}{4}(a+b-c)^2+\frac{1}{4}}. 
\end{array}
\end{equation}
Then, we have 
\begin{equation}
\tau_i=t^{A_i}(t-1)^{B_i}, \quad s_2s_i(\tau_i)=0 \quad (i=0,1,3,4), 
\end{equation}
\begin{equation}
\begin{array}{ll}
\medskip
{\disp s_0(\tau_0)=b\varphi_{-1,-2,-1}t^{A_0+b}(t-1)^{B_0},} & 
{\disp s_1(\tau_1)=\varphi_{0,-1,0}t^{A_1+(c-a)}(t-1)^{B_1+(a+b-c+1)},} \\
{\disp s_3(\tau_3)=(c-a)\varphi_{-1,-1,0}t^{A_3+a}(t-1)^{B_3},} & 
{\disp s_4(\tau_4)=\varphi_{-1,-1,-1}t^{A_4-(b-c+1)}(t-1)^{B_4+(a+b-c+1)},}
\end{array}
\end{equation}
up to multiplication by some constants. 
For small $k,l,m,n$, we observe that $\tau_{k,l,m,n}$ are expressed in
the form 
\begin{equation}
\tau_{k,l,m,n}=\sigma_{k,l,m,n}
t^{-\frac{1}{4}(\hat{a}+\hat{b}-\hat{c}+2n)^2-\frac{1}{4}(\hat{a}-\hat{b}-n)^2+n(\hat{b}+n)-\frac{1}{2}n(n-1)}
(t-1)^{\frac{1}{4}(\hat{a}+\hat{b}-\hat{c}+2n)^2+\frac{1}{2}}, 
\label{tau-sig:P6}
\end{equation}
with 
\begin{equation}
\sigma_{k,l,m,-1}=0, \quad 
\sigma_{k,l,m,0}=\mbox{const.}, \quad 
\sigma_{k,l,m,1}=(\mbox{const.})\times\varphi_{k,l,m}, 
\end{equation}
where we denote 
\begin{equation}
\hat{a}=a+k, \quad \hat{b}=b+l+1, \quad \hat{c}=c+m. 
\end{equation}
Assume that $\tau_{k,l,m,n}$ are expressed as (\ref{tau-sig:P6}) for any
$k,l,m,n\in\BZ$. 
Then, the Toda equation (\ref{Toda:P6}) yields 
\begin{equation}
\sigma_{k,l,m,n+1}\sigma_{k,l,m,n-1}=
-\left[(\dl^2\sigma_{k,l,m,n})\sigma_{k,l,m,n}-(\dl\sigma_{k,l,m,n})^2\right], 
\quad \dl=t\frac{d}{dt}. 
\end{equation}
Moreover, we set 
\begin{equation}
\sigma_{k,l,m,n}=\omega_{k,l,m,n}\rho_{k,l,m,n}, \quad 
\omega_{k,l,m,n}=\omega_{k,l,m,n}(a,b,c), 
\end{equation}
with $\rho_{k,l,m,0}=1$ and $\rho_{k,l,m,1}=\varphi_{k,l,m}$, 
and impose that the constants $\omega_{k,l,m,n}$ satisfy 
\begin{equation}
\omega_{k,l,m,n+1}\omega_{k,l,m,n-1}=-\omega_{k,l,m,n}^2. \label{omega:k:P6}
\end{equation}
Then, the function $\rho_{k,l,m,n}$ are determined by the recurrence
relation 
\begin{equation}
\rho_{k,l,m,n+1}\rho_{k,l,m,n-1}=
(\dl^2\rho_{k,l,m,n})\rho_{k,l,m,n}-(\dl\rho_{k,l,m,n})^2, 
\end{equation}
with the initial conditions 
\begin{equation}
\rho_{k,l,m,-1}=0, \quad \rho_{k,l,m,0}=1, \quad \rho_{k,l,m,1}=\varphi_{k,l,m}. 
\end{equation}
By Darboux's formula, the functions $\rho_{k,l,m,n}$ for 
$n\in\BZ_{\ge 0}$ are expressed as 
\begin{equation}
\rho_{k,l,m,n}=
 \left|
 \begin{array}{cccc}
  \varphi_{k,l,m}^{(0)}   & \varphi_{k,l,m}^{(1)} & \cdots & \varphi_{k,l,m}^{(n-1)}  \\
  \varphi_{k,l,m}^{(1)}   & \varphi_{k,l,m}^{(2)} & \cdots & \varphi_{k,l,m}^{(n)}    \\
  \vdots                  & \vdots                & \ddots & \vdots                   \\
  \varphi_{k,l,m}^{(n-1)} & \varphi_{k,l,m}^{(n)} & \cdots & \varphi_{k,l,m}^{(2n-2)}
 \end{array}
 \right|,   \quad \varphi_{k,l,m}^{(i)}=\left(t\frac{d}{dt}\right)^i\varphi_{k,l,m}. 
\end{equation}
Note that the constants $\omega_{k,l,m,n}$ are determined by the
recurrence relations (\ref{omega:k:P6}) and 
\begin{equation}
\begin{array}{l}
\smallskip
\disp
\omega_{k+1,l,m,i}\omega_{k-1,l,m,i}=i\hat{a}(\hat{c}-\hat{a})\omega_{k,l,m,i}^2, \\
\smallskip
\disp
\omega_{k,l+1,m,i}\omega_{k,l-1,m,i}=-i\hat{b}(\hat{c}-\hat{b})\omega_{k,l,m,i}^2, \\
\disp
\omega_{k,l,m+1,i}\omega_{k,l,m-1,i}=(\hat{c}-\hat{a})(\hat{c}-\hat{b})\omega_{k,l,m,i}^2, 
\end{array}   \quad (i=0,1)
\end{equation}
with the initial conditions 
\begin{equation}
\begin{array}{l}
\smallskip
\disp
\omega_{-1,-2,-1,1}=(-1)^{-\frac{1}{4}}b, \quad 
\omega_{ 0,-2,-1,1}=b, \quad 
\omega_{-1,-1,-1,1}=1, \quad 
\omega_{ 0,-1,-1,1}=(-1)^{-\frac{1}{4}}, \\
\disp
\omega_{-1,0,0,1}=-(-1)^{-\frac{3}{4}}(c-a), \quad 
\omega_{ 0,0,0,1}=-i, \quad 
\omega_{-1,-1,0,1}=-i(c-a), \quad 
\omega_{ 0,-1,0,1}=(-1)^{-\frac{3}{4}}, 
\end{array}
\end{equation}
and 
\begin{equation}
\begin{array}{c}
\smallskip
\disp
\omega_{-1,-2,-1,0}=(-1)^{-\frac{3}{4}}b, \quad 
\omega_{ 0,-2,-1,0}=-b, \quad 
\omega_{-1,-1,-1,0}=1, \quad 
\omega_{ 0,-1,-1,0}=(-1)^{-\frac{3}{4}}, \\
\disp
\omega_{-1,0,0,0}=(-1)^{-\frac{3}{4}}(c-a), \quad 
\omega_{ 0,0,0,0}=1, \quad 
\omega_{-1,-1,0,0}=c-a, \quad 
\omega_{ 0,-1,0,0}=(-1)^{-\frac{3}{4}}. 
\end{array}
\end{equation}
Since it is possible to set $k=l=m=0$ without loss of generality, 
we obtain the following theorem. 

\begin{thm}\label{P6:spec}
We define the functions $\tau_n^{k,l,m}$ by 
\begin{equation}
\tau_n^{k,l,m}=
 \left|
 \begin{array}{cccc}
  \varphi_{k,l,m}^{(0)}   & \varphi_{k,l,m}^{(1)} & \cdots & \varphi_{k,l,m}^{(n-1)}  \\
  \varphi_{k,l,m}^{(1)}   & \varphi_{k,l,m}^{(2)} & \cdots & \varphi_{k,l,m}^{(n)}    \\
  \vdots                  & \vdots                & \ddots & \vdots                   \\
  \varphi_{k,l,m}^{(n-1)} & \varphi_{k,l,m}^{(n)} & \cdots & \varphi_{k,l,m}^{(2n-2)}
 \end{array}
 \right|,   \quad \varphi_{k,l,m}^{(i)}=\left(t\frac{d}{dt}\right)^i\varphi_{k,l,m}, 
\end{equation}
where $\varphi_{k,l,m}$ are given by (\ref{entry:6}). 
Then, 
\begin{equation}
\begin{array}{l}
\medskip
\disp
f_0=b
    \frac{\tau_n^{0,0,0}\tau_{n+1}^{-1,-2,-1}}
         {\tau_n^{-1,-1,-1}\tau_{n+1}^{0,-1,0}}, \quad 
f_3=(c-a)
    \frac{\tau_n^{0,-1,-1}\tau_{n+1}^{-1,-1,0}}
         {\tau_n^{-1,-1,-1}\tau_{n+1}^{0,-1,0}}, \\
\disp
f_4=\frac{\tau_n^{0,-1,0}\tau_{n+1}^{-1,-1,-1}}
         {\tau_n^{-1,-1,-1}\tau_{n+1}^{0,-1,0}}, \quad 
f_2=at^{-1}
    \frac{\tau_n^{-1,-1,-1}\tau_{n+1}^{0,-1,0}\tau_{n-1}^{1,0,0}}
         {\tau_n^{0,0,0}\tau_n^{0,-1,-1}\tau_n^{0,-1,0}}, 
\end{array}
\end{equation}
\begin{equation}
(\aa_0,\aa_1,\aa_2,\aa_3,\aa_4)=(-b,a+n,-n,c-a,b-c+1+n), 
\end{equation}
gives a family of classical transcendental solutions of the symmetric 
form of P$_{\rm VI}$. 
\end{thm}

\section{Degeneration of classical transcendental solutions\label{deg}}
It is well known that, starting from P$_{\rm VI}$, one can obtain
P$_{\rm V}, \ldots$, P$_{\rm I}$ by successive limiting procedures in
the following diagram~\cite{Pa,GtoP}, 
\begin{equation}
\begin{array}{c}
\mbox{P}_{\rm VI} \lora \mbox{P}_{\rm V}  \lora \mbox{P}_{\rm III} \\
\hskip30pt                \da        \hskip30pt   \da              \\
\hskip71pt              \mbox{P}_{\rm IV} \lora \mbox{P}_{\rm II}  \lora \mbox{P}_{\rm I}, 
\end{array}
\end{equation}
which corresponds to the degeneration diagram of the special functions
of hypergeometric type 
\begin{equation}
\begin{array}{c}
\mbox{Gauss}\lora\mbox{Kummer}\lora\mbox{Bessel} \\
\hskip60pt      \da  \hskip50pt   \da \\
\hskip35pt      \mbox{Hermite-Weber}\lora\mbox{Airy}. 
\end{array}
\end{equation}
In this section, we show that, starting from the family of special
function solutions of P$_{\rm VI}$ given in Theorem \ref{P6:spec}, 
we obtain classical transcendental solutions to other Painlev\'e
equations by degeneration.

\subsection{From P$_{\rm VI}$ to P$_{\rm V}$}
As is known, the Hamiltonian system S$_{\rm VI}$ is reduced to 
S$_{\rm V}$ by putting 
\begin{equation}
t \to 1-\ep t, \quad \kappa_1 \to \ep^{-1}+\theta+1, \quad \theta \to -\ep^{-1}, 
\end{equation}
and taking the limit as $\ep\to 0$. 
Let us consider the degeneration of the family of classical
transcendental solutions given in Theorem \ref{P6:spec}. 
It is known that P$_{\rm VI}$ admits an outer symmetry as 
\begin{equation}
\sigma_{34}: \quad \aa_3 \leftrightarrow \aa_4, \quad t \ra 1-t, \quad 
             f_4 \ra  -f_3, \quad f_2 \ra  -f_2. 
\end{equation}
The application of $\sigma_{34}$ to the family of solutions in Theorem 
\ref{P6:spec} gives another family of solutions for the parameters 
\begin{equation}
(\aa_0,\aa_1,\aa_2,\aa_3,\aa_4)=(-b,a+n,-n,b-c+1+n,c-a), 
\end{equation}
or 
\begin{equation}
\ka_{\infty}=a+n, \quad \ka_0=c-a, \quad \ka_1=b-c+1+n, \quad \theta=-b. \label{para':P6}
\end{equation}
Then, it is easy to see that by putting 
\begin{equation}
t \to \ep t, \quad b=\ep^{-1}, 
\end{equation}
S$_{\rm VI}$ with (\ref{para':P6}) is reduced to S$_{\rm V}$ with 
\begin{equation}
\ka_{\infty}=a+n, \quad \ka_0=c-a, \quad \theta=-c+n, 
\end{equation}
in the limit as $\ep\to 0$. 
It is obvious that Gauss's function $F(a,b,c;t)$ is reduced to Kummer's
function $F(a,c;t)$ by this process. 
Thus, we get 
\begin{equation}
\varphi_{k,l,m}\to\ep^{-l}\Gm^{-1}\ms{$(c-a-k+m)$}\varphi_{k+1,m+1}, 
\end{equation}
where $\varphi_{i,j}$ are given by (\ref{entry:5}). 
Note that we redefine the constants $c_1$ and $c_2$ appropriately. 
It is easy to see that we have 
\begin{equation}
\tau^{k,l,m}_n\to\ep^{-ln}\Gm^{-n}\ms{$(c-a-k+m)$}\tau_n^{k+1,m+1}. 
\end{equation}
Therefore, we obtain the family of classical transcendental solutions of
S$_{\rm V}$ in Theorem \ref{P5:spec}.

\subsection{From P$_{\rm V}$ to P$_{\rm III}$}
From P$_{\rm V}$, we can obtain two coalescence limits. 
First, we consider the Painlev\'e III equation 
\begin{equation}
\frac{d^2q}{dt^2}=
\frac{1}{q}\left(\frac{dq}{dt}\right)^2-\frac{1}{t}\frac{dq}{dt}
-\frac{4}{t}\left[\eta_{\infty}\theta_{\infty}q^2+\eta_0(\theta_0+1)\right]
+4\eta_{\infty}^2 q^3-\frac{4\eta_0^2}{q}, 
\end{equation}
which is equivalent to the Hamiltonian system 
\begin{equation}
\hskip-40pt
\mbox{S$_{\rm III}$~:} \hskip30pt
q'=\frac{\partial H}{\partial p}, \quad p'=-\frac{\partial H}{\partial q}, \quad
'=t\frac{d}{dt}, 
\end{equation}
with the Hamiltonian 
\begin{equation}
H=2q^2p^2-\left[2\eta_{\infty}tq^2+(2\theta_0+1)q+2\eta_0 t\right]p+\eta_{\infty}(\theta_0+\theta_{\infty})tq. 
\end{equation}
This system can be derived from S$_{\rm V}$ by putting 
\begin{equation}
\begin{array}{c}
\medskip
\disp
q\to 1+\ep tq, \quad p\to\ep^{-1}t^{-1}p, \quad 
t\to\eta_0\ep t^2, \quad H\to\frac{1}{2}(H+qp)\\
\disp
\ka_{\infty}\to -\eta_{\infty}\ep^{-1}+\theta_{\infty}, \quad 
\ka_0\to\eta_{\infty}\ep^{-1}, \quad \theta \to \theta_0, 
\end{array}
\end{equation}
and taking the limit as $\ep \to 0$. 

Let us apply the limiting procedure to the family of classical
transcendental solutions of S$_{\rm V}$ given in Theorem \ref{P5:spec}. 
It is easy to see that by putting 
\begin{equation}
a \to -\eta_{\infty}\ep^{-1}+c, 
\end{equation}
S$_{\rm V}$ with (\ref{para:P5}) is reduced to S$_{\rm III}$ with 
\begin{equation}
\theta_{\infty}=\nu+n+1, \quad \theta_0=-\nu+n-1, 
\end{equation}
in the limit as $\ep\to 0$, where we denote $c=\nu+1$. 
Without loss of generality, it is possible to set 
\begin{equation}
\eta_{\infty}=\frac{\sigma}{2}, \quad \eta_0=\frac{1}{2}, \quad (\sigma=\pm1). 
\label{fix}
\end{equation}
Then, we find that 
\begin{equation}
F(a,c;t)\to\Gamma(\nu+1)\left(\frac{t}{2}\right)^{-\nu}Z_{\nu}(t), 
\end{equation}
where  $Z_{\nu}=Z_{\nu}(t)$ denotes 
\begin{equation}
Z_{\nu} = 
 \left\{
  \begin{array}{ccc}
   \medskip
   J_{\nu}: & \mbox{Bessel}          & \sigma=+1, \\
   I_{\nu}: & \mbox{modified Bessel} & \sigma=-1. 
  \end{array}
 \right.
\end{equation}
This means that Kummer's function is reduced to the (modified) Bessel
function in this limit. 
Thus, we get 
\begin{equation}
\varphi_{i,j}\to(-1)^i\left(\frac{\sigma}{2}\ep^{-1}\right)^j
                \left(\frac{t}{2}\right)^{-\nu-j}\varphi_{\nu+j}, 
\end{equation}
with 
\begin{equation}
\varphi_{\nu+j}=c_1Z_{\nu+j}+c_2(-\sigma)^jZ_{-\nu-j}. \label{phi:P3}
\end{equation}
This leads us to 
\begin{equation}
\tau_n^{i,j}\to
\left(\frac{1}{2}\right)^{(n-1)n}(-1)^{in}
\left(\frac{\sigma}{2}\ep^{-1}\right)^{jn}
\left(\frac{t}{2}\right)^{-(\nu+j)n}\tau_n^{\nu+j}, 
\end{equation}
where $\tau_n^{\nu+j}$ are defined by 
\begin{equation}
\tau_n^{\nu+j}=
 \left|
 \begin{array}{cccc}
  \varphi_{\nu+j}^{(0)}   & \varphi_{\nu+j}^{(1)} & \cdots & \varphi_{\nu+j}^{(n-1)} \\
  \varphi_{\nu+j}^{(1)}   & \varphi_{\nu+j}^{(2)} & \cdots & \varphi_{\nu+j}^{(n)}   \\
  \vdots                  & \vdots                & \ddots & \vdots                  \\
  \varphi_{\nu+j}^{(n-1)} & \varphi_{\nu+j}^{(n)} & \cdots & \varphi_{\nu+j}^{(2n-2)}
 \end{array}
 \right|,   \quad 
\varphi_{\nu+j}^{(i)}=\left(t\frac{d}{dt}\right)^i\varphi_{\nu+j}. \label{tau:P3}
\end{equation}
Therefore, from Theorem \ref{P5:spec} and Remark \ref{rem:P5}, 
we obtain the following proposition~\cite{O4}. 

\begin{prop}~(Okamoto)
We define the functions $\tau_n^{\nu}$ by (\ref{tau:P3}). 
Then, 
\begin{equation}
q=\sigma\frac{\tau_{n+1}^{\nu}\tau_n^{\nu+1}}{\tau_n^{\nu}\tau_{n+1}^{\nu+1}}
 =\sigma\left(\frac{d}{dt}\log\frac{\tau_{n+1}^{\nu+1}}{\tau_n^{\nu}}
              +\frac{\nu+1-n}{t}\right), \quad 
p=-\frac{1}{2t}\frac{\tau_{n+1}^{\nu+1}\tau_{n-1}^{\nu+1}}{\tau_n^{\nu+1}\tau_n^{\nu+1}},
\end{equation}
\begin{equation}
\theta_{\infty}=\nu+n+1, \quad \theta_0=-\nu+n-1, \label{para:P3}
\end{equation}
with (\ref{fix}) gives a family of classical transcendental solutions of
S$_{\rm III}$. 
\end{prop}

\subsection{From P$_{\rm V}$ to P$_{\rm IV}$}
Next, we consider the Painlev\'e IV equation 
\begin{equation}
\frac{d^2q}{dt^2}=
\frac{1}{2q}\left(\frac{dq}{dt}\right)^2+\frac{3}{2}q^3+2tq^2
+\frac{1}{2}t^2q-(-\ka_0+2\theta_{\infty}+1)q-\frac{\ka_0^2}{2q},
\end{equation}
which is equivalent to the Hamiltonian system 
\begin{equation}
\hskip-40pt
\mbox{S$_{\rm IV}$~:} \hskip30pt
q'=\frac{\partial H}{\partial p}, \quad p'=-\frac{\partial H}{\partial q}, \quad
'=\frac{d}{dt}, 
\end{equation}
with the Hamiltonian 
\begin{equation}
H=qp^2-(q^2+tq+\ka_0)p+\theta_{\infty}q. 
\end{equation}
This system can be also derived from S$_{\rm V}$ by coalescence. 
This process is achieved by putting 
\begin{equation}
\begin{array}{c}
\medskip
\disp
q \to \ep q, \quad p \to \ep^{-1}p, \quad 
t \to \ep^{-2}(1+\ep t), \quad H+\ka \to \ep^{-1}H, \\
\disp
\theta \to \ep^{-2}+2\theta_{\infty}-\ka_0, \quad 
\ka_{\infty} \to \ep^{-2}, 
\end{array}
\end{equation}
and taking the limit as $\ep \to 0$. 

Let us consider the degeneration of the classical transcendental
solutions of S$_{\rm V}$. 
Applying the B\"acklund transformation $\pi^2$ to the solutions in
Theorem \ref{P5:spec}, we obtain the following corollary. 

\begin{coro}
We define the functions $\tau_n^{i,j}$ by (\ref{tau:P5}). 
Then, 
\begin{equation}
q=-\left(\frac{c-a}{c-a-1}\right)^n
\frac{\tau_n^{0,0}\tau_{n+1}^{1,1}}
     {\tau_n^{1,0}\tau_{n+1}^{0,1}}, \quad 
p=(1-a)\left(\frac{c-a-1}{c-a}\right)^n
  \frac{\tau_n^{1,0}\tau_{n+1}^{0,1}\tau_{n+1}^{-1,-1}}
       {\tau_{n+1}^{0,0}\tau_{n+1}^{0,0}\tau_n^{0,0}}, 
\end{equation}
\begin{equation}
\ka_{\infty}=c-a, \quad \ka_0=a+n, \quad \theta=c-n-2, \label{para':P5}
\end{equation}
gives a family of classical transcendental solutions of S$_{\rm V}$. 
\end{coro}

It is easy to see that by putting 
\begin{equation}
c \to \ep^{-2}+a, 
\end{equation}
S$_{\rm V}$ with (\ref{para':P5}) is reduced to S$_{\rm IV}$ with 
\begin{equation}
\ka_0=-\nu+n, \quad \theta_{\infty}=-\nu-1, 
\end{equation}
in the limit as $\ep\to 0$, where we denote $a=-\nu$. 
We find that 
\begin{equation}
\frac{\Gm(c-a)}{\Gm(c)}F(a,c;t)\to(-\ep)^{-\nu}H_{\nu}(t), 
\end{equation}
where $H_{\nu}(t)$ denotes the Hermite-Weber function. 
By a Kummer transformation
\begin{equation}
F(a-c+1,2-c;t)=e^tF(1-a,2-c;-t), 
\end{equation}
we get 
\begin{equation}
\frac{\Gm(c-a)\Gm(a-c+1)}{\Gm(2-c)}t^{1-c}F(a-c+1,2-c;t)\to
(-i\ep)^{-\nu-1}e^{\frac{t^2}{2}}H_{-\nu-1}(it). 
\end{equation}
Thus, we have 
\begin{equation}
\varphi_{k,j}\to(-\ep)^{-\nu+k}
\left[c_1\Gm(-\nu+k)H_{\nu-k}(t)
     +c_2e^{i\pi(-\nu+k-1)/2}e^{\frac{t^2}{2}}H_{-\nu+k-1}(it)\right]. 
\end{equation}
Let us rewrite this expression in terms of the hyperbolic cylinder
function $D_{\nu}(t)$. 
Noting the relations 
\begin{equation}
H_{\nu}(t)=e^{\frac{t^2}{4}}D_{\nu}(t), 
\end{equation}
and 
\begin{equation}
D_{-\nu-1}(it)=
\frac{\Gamma(-\nu)}{\sqrt{2\pi}}\left[
e^{i\pi(\nu+1)/2}D_{\nu}(t)-e^{-i\pi(\nu-1)/2}D_{\nu}(-t)\right], 
\end{equation}
we get 
\begin{equation}
\varphi_{k,j}\to \ep^ke^{\frac{t^2}{4}}\varphi_{\nu-k}, 
\end{equation}
with 
\begin{equation}
\varphi_{\nu-k}=c_1\frac{D_{\nu-k}(t)}{\Gm(\nu-k+1)}+c_2\Gm(-\nu+k)D_{\nu-k}(-t). 
\label{phi:P4}
\end{equation}
This leads us to 
\begin{equation}
\tau_n^{k,j}\to\ep^{-n(n-1)}\ep^{kn}\tau_n^{\nu-k}, 
\end{equation}
where $\tau_n^{\nu-k}$ are defined by 
\begin{equation}
\tau_n^{\nu-k}=
 \left|
 \begin{array}{cccc}
  \phi_{\nu-k}^{(0)}   & \phi_{\nu-k}^{(1)} & \cdots & \phi_{\nu-k}^{(n-1)}  \\
  \phi_{\nu-k}^{(1)}   & \phi_{\nu-k}^{(2)} & \cdots & \phi_{\nu-k}^{(n)}    \\
  \vdots               & \vdots             & \ddots & \vdots                \\
  \phi_{\nu-k}^{(n-1)} & \phi_{\nu-k}^{(n)} & \cdots & \phi_{\nu-k}^{(2n-2)}
 \end{array}
 \right|,   \quad 
\phi_{\nu-k}^{(m)}=
\left(\frac{d}{dt}\right)^m\left(e^{\frac{t^2}{4}}\varphi_{\nu-k}\right). 
\label{tau:P4}
\end{equation}
Therefore, we obtain the following proposition. 

\begin{prop}\label{spec:P4'}
We define the functions $\tau_n^{\nu-k}$ by (\ref{tau:P4}). 
Then, 
\begin{equation}
q=-\frac{\tau_n^{\nu}\tau_{n+1}^{\nu-1}}
        {\tau_n^{\nu-1}\tau_{n+1}^{\nu}}, \quad 
p=(\nu+1)
  \frac{\tau_n^{\nu-1}\tau_{n+1}^{\nu+1}}
       {\tau_{n+1}^{\nu}\tau_n^{\nu}}, 
\end{equation}
\begin{equation}
\ka_0=-\nu+n, \quad \theta_{\infty}=-\nu-1, 
\end{equation}
gives a family of classical transcendental solutions of S$_{\rm IV}$. 
\end{prop}

The direct derivation of this proposition is given in Appendix. 

\begin{rem}
A special case ($c_2=0$) of Proposition \ref{spec:P4'} is stated in
\cite{P4:rat}. 
In \cite{NY:P4}, the case of $\nu\in\BZ$, where the $\tau$-functions are 
reduced to some polynomials, is discussed. 
\end{rem}

\subsection{From P$_{\rm III}$ to P$_{\rm II}$}
Both P$_{\rm III}$ and P$_{\rm IV}$ go to P$_{\rm II}$ by coalescence. 
First, we consider the degeneration from $S_{\rm III}$ to $S_{\rm II}$, 
which is achieved by putting 
\begin{equation}
\begin{array}{c}
\medskip
\disp
q \to 1+\ep q, \quad p \to \ep^{-1}p, \quad 
\theta_0 \to 2\ep^{-3}+\aa^{(1)}, \quad \theta_{\infty} \to -2\ep^{-3}+\aa^{(2)}, \\
\disp
t \to -2\ep^{-3}\left(1-\frac{1}{2}\ep^2t\right), \quad 
H \to -2\ep^{-2}H-2\ep^{-3}\aa, 
\end{array}   \label{3to2}
\end{equation}
and taking the limit as $\ep \to 0$, where we set
$\aa=\frac{1}{2}\left(\aa^{(1)}+\aa^{(2)}\right)$. 

Let us consider the degeneration of the family of classical
transcendental solutions. 
From (\ref{3to2}), it is the case of $\sigma=1$ (the Bessel function)
that we can take the degeneration limit. 
By the relation 
\begin{equation}
J_{-\nu}=\cos(\nu\pi)J_{\nu}-\sin(\nu\pi)Y_{\nu}, 
\end{equation}
we rewrite $\varphi_{\nu+j}$ (\ref{phi:P3}) as 
\begin{equation}
\varphi_{\nu+j}=c_1J_{\nu+j}+c_2Y_{\nu+j}. 
\end{equation}
Then, from (\ref{3to2}), we see that by putting 
\begin{equation}
\nu=-2\ep^{-3}, 
\end{equation}
S$_{\rm III}$ with (\ref{para:P3}) is reduced to S$_{\rm II}$ with
$\aa=n$. 
It is known that we have~\cite{AS} 
\begin{equation}
\begin{array}{l}
\medskip
\disp
J_{\nu}(\nu+z\nu^{\frac{1}{3}})=
2^{\frac{1}{3}}\nu^{-\frac{1}{3}}\Ai(-2^{\frac{1}{3}}z)+O(\nu^{-1}), \\
\disp
Y_{\nu}(\nu+z\nu^{\frac{1}{3}})=
-2^{\frac{1}{3}}\nu^{-\frac{1}{3}}\Bi(-2^{\frac{1}{3}}z)+O(\nu^{-1}), 
\end{array}
\end{equation}
which lead us to 
\begin{equation}
J_{\nu}(t)\to -\ep\Ai(t)+O(\ep^3), \quad Y_{\nu}(t)\to \ep\Bi(t)+O(\ep^3). 
\end{equation}
Thus, we get 
\begin{equation}
\tau_n^{\nu+j}\to\ep^n(-2\ep^{-2})^{(n-1)n}\tau_n, 
\end{equation}
where the functions $\tau_n$ are defined by (\ref{tau:P2}). 
Therefore, we obtain Proposition \ref{spec:P2}.

\subsection{From P$_{\rm IV}$ to P$_{\rm II}$}
It is well known that the Hamiltonian system S$_{\rm II}$ is also
derived from S$_{\rm IV}$ by degeneration. 
This process is achieved by putting 
\begin{equation}
\begin{array}{c}
\medskip
\disp
q\to\ep^{-3}(1-\ep^2q), \quad p\to -\ep p, \quad 
t\to -2\ep^{-3}\left(1+\frac{1}{2}\ep^4t\right), \\
\disp
\ka_0\to\ep^{-6}, \quad \theta_{\infty}\to \aa, \quad H\to -\ep^{-1}H+\ep^{-3}\aa,
\end{array}
\end{equation}
and taking the limit as $\ep\to 0$. 

Let us consider the degeneration of the classical transcendental
solutions. 
Applying the B\"acklund transformation $\pi$ to the solutions in Theorem
\ref{spec:P4}, we obtain the following corollary. 

\begin{coro}
We define the functions $\tau_n^{\nu}$ by (\ref{tau:P4}). 
Then, 
\begin{equation}
q=\frac{d}{dt}\log\frac{\tau_{n+1}^{\nu}}{\tau_n^{\nu}}-t, \quad 
p=\frac{\tau_{n+1}^{\nu}\tau_{n-1}^{\nu-1}}
         {\tau_n^{\nu-1}\tau_n^{\nu}}, 
\end{equation}
\begin{equation}
\ka_0=\nu+1, \quad \theta_{\infty}=n, \label{para:P4}
\end{equation}
gives a family of classical transcendental solutions of S$_{\rm IV}$. 
\end{coro}

It is easy to see that by putting 
\begin{equation}
\nu\to\ep^{-6}-1, 
\end{equation}
S$_{\rm IV}$ with (\ref{para:P4}) reduced to S$_{\rm II}$ with $\aa=n$
in the limit as $\ep\to 0$. 

Let us consider the degeneration of classical transcendental solutions. 
According to \cite{Ha}, we find that the parabolic cylinder function is 
reduced to the Airy function as 
\begin{equation}
\frac{D_{\nu+j}(t)}{\Gm(\nu+j+1)}\to(-\ep)^{3j}\Ai(t), \quad 
\Gm(-\nu-j)D_{\nu+j}(-t)\to(-\ep)^{3j}\Ai(\omega t), 
\end{equation}
with $\omega=e^{2\pi i/3}$. 
Thus, we have 
\begin{equation}
\varphi_{\nu+j}\to(-\ep)^{3j}\varphi, 
\end{equation}
where $\varphi$ denote the general solution of Airy's differential
equation. 
Normalizing the $\tau$-functions (\ref{tau:P4}) as 
\begin{equation}
\tau_n^{\nu+j}=e^{n\frac{t^2}{4}}\widetilde{\tau}_n^{\nu+j}, 
\end{equation}
we get 
\begin{equation}
\widetilde{\tau}_n^{\nu+j}\to(-\ep)^{-n(n-1)+3jn}\tau_n, 
\end{equation}
where the functions $\tau_n$ are defined by (\ref{tau:P2}). 
Therefore, we obtain Proposition \ref{spec:P2}. 

\medskip

\noindent 
{\bf Acknowledgment}\quad 
The author would like to thank Prof. Y. Haraoka, Prof. K. Kajiwara and
Prof. M. Noumi for their helpful comments.

\appendix
\section{Classical transcendental solutions of the Painlev\'e IV equation}
\subsection{The symmetric form of the Painlev\'e IV equation}
The symmetric form of P$_{\rm IV}$ is given by~\cite{NY:P4}
\begin{equation}
\begin{array}{l}
\medskip
\disp 
 f_0'=f_0(f_1-f_2)+\aa_0, \\
\medskip
\disp 
 f_1'=f_1(f_2-f_0)+\aa_1, \\
\disp 
 f_2'=f_2(f_0-f_1)+\aa_2,
\end{array}   \quad '=\frac{d}{dt}, \label{sym:A2}
\end{equation}
with the normalization conditions $\aa_0+\aa_1+\aa_2=1$ and
$f_0+f_1+f_2=t$. 
The correspondence to the canonical variables and parameters of S$_{\rm
IV}$ is given by 
\begin{equation}
q=-f_1, \quad p=f_2, \qquad 
\ka_0=\aa_1, \quad \theta_{\infty}=-\aa_2. 
\end{equation}
The B\"acklund transformations of P$_{\rm IV}$ are described as follows, 
\begin{equation}
\begin{array}{lll}
\medskip
{\disp s_i(\aa_i)=-\aa_i,} & {\disp s_i(\aa_j)=\aa_j+\aa_i~~(j=i \pm 1),} & 
{\disp \pi(\aa_j)=\aa_{j+1},} \\ 
{\disp s_i(f_i)=f_i,} & {\disp s_i(f_j)=f_j \pm \frac{\aa_i}{f_i}~~(j=i \pm 1),} &
{\disp \pi(f_j)=f_{j+1},}
\end{array}
\end{equation}
where the subscripts $i=0,1,2$ are understood as elements of
$\BZ/3\BZ$. 
The Hamiltonians $h_i$ of the system (\ref{sym:A2}) are given by 
\begin{equation}
h_0=f_0f_1f_2
    +\frac{\aa_1-\aa_2}{3}f_0+\frac{\aa_1+2\aa_2}{3}f_1-\frac{2\aa_1+\aa_2}{3}f_2, 
\end{equation}
and $h_i=\pi^i(h_0)$. 
Introducing the $\tau$-functions $\tau_i$ as $h_i=(\log\tau_i)'$, 
we find that the B\"acklund transformations for the $\tau$-functions are
described by 
\begin{equation}
s_i(\tau_j)=\tau_j~~(i\neq j), \quad s_i(\tau_i)=f_i\frac{\tau_{i-1}\tau_{i+1}}{\tau_i}, 
\quad \pi(\tau_i)=\tau_{i+1}. 
\end{equation}
The $f$-variables are recovered from the $\tau$-functions by 
\begin{equation}
f_i=\frac{\tau_is_i(\tau_i)}{\tau_{i-1}\tau_{i+1}}
   =\frac{d}{dt}\log\frac{\tau_{i-1}}{\tau_{i+1}}+\frac{t}{3}. \label{f-tau}
\end{equation}

Let us define the translation operators $T_i$ by 
\begin{equation}
T_1=\pi s_2s_1, \quad T_2=s_1\pi s_2, \quad T_0=s_2s_1\pi, 
\end{equation}
which commute with each other and act on the parameters $\aa_i$ as 
\begin{equation}
T_i(\aa_{i-1})=\aa_{i-1}+1, \quad T_i(\aa_i)=\aa_i-1, \quad 
T_i(\aa_j)=\aa_j~(j \ne i-1,i).
\end{equation}
Noting that $T_1T_2T_0=1$, we set 
\begin{equation}
\tau_{k,l}=T_1^kT_2^l(\tau_0), \quad k,l \in \BZ. 
\end{equation}
Then, we have from (\ref{f-tau}) 
\begin{equation}
\begin{array}{l}
\medskip
\disp
T_1^kT_2^l(f_0)
=\frac{\tau_{k,l}\tau_{k+2,l+1}}
      {\tau_{k+1,l+1}\tau_{k+1,l}}
=\frac{d}{dt}\log\frac{\tau_{k+1,l+1}}{\tau_{k+1,l}}+\frac{t}{3}, \\
\medskip
\disp
T_1^kT_2^l(f_1)
=\frac{\tau_{k+1,l}\tau_{k,l+1}}
      {\tau_{k,l}\tau_{k+1,l+1}}
=\frac{d}{dt}\log\frac{\tau_{k,l}}{\tau_{k+1,l+1}}+\frac{t}{3}, \\
\disp
T_1^kT_2^l(f_2)
=\frac{\tau_{k+1,l+1}\tau_{k,l-1}}
      {\tau_{k+1,l}\tau_{k,l}}
=\frac{d}{dt}\log\frac{\tau_{k+1,l}}{\tau_{k,l}}+\frac{t}{3}. 
\end{array}
\end{equation}
It is possible to derive the Toda equation with respect to each
translation operator. 
For the $T_1$-direction, we have 
\begin{equation}
\tau_{k+1,l}\tau_{k-1,l}=
\left[(\log\tau_{k,l})''+\frac{2\aa_1+\aa_2-2k+l}{3}\right]\tau_{k,l}\cdot\tau_{k,l}. 
\label{Toda:P4}
\end{equation}

\subsection{A Riccati solution}
We derive a Riccati solution of (\ref{sym:A2}). 
First, we set $\aa_0=0$ and $f_0=0$. 
Then, $f_1$ satisfies a Riccati equation 
\begin{equation}
f_1'=f_1(t-f_1)+\aa_1. 
\end{equation}
By the dependent variable transformation
$f_1=(\log\varphi)'+\frac{t}{2}$, 
we have for $\varphi$ the linear equation 
\begin{equation}
\left(\frac{d^2}{dt^2}-\aa_1+\frac{1}{2}-\frac{t^2}{4}\right)\varphi=0, \label{HW}
\end{equation}
which is nothing but Weber's differential equation. 
We set $\aa_1=-\nu~(\not\in\BZ)$. 
Then, the general solution of (\ref{HW}) is expressed by 
\begin{equation}
\varphi=c_1\frac{D_{\nu}(t)}{\Gm(\nu+1)}+c_2\Gm(-\nu)D_{\nu}(-t), 
\end{equation}
where $c_i~(i=1,2)$ are arbitrary complex constants. 
By using the contiguity relations of the hyperbolic cylinder function, 
we obtain the following proposition. 

\begin{prop}\label{P4:Ric}
We define the function $\varphi_{\nu-k}$ by (\ref{phi:P4}). 
Then, 
\begin{equation}
(f_0,f_1,f_2)=
\left(0,\frac{\varphi_{\nu-1}}{\varphi_{\nu}},
      (\nu+1)\frac{\varphi_{\nu+1}}{\varphi_{\nu}}\right), 
\quad 
(\aa_0,\aa_1,\aa_2)=\left(0,-\nu,\nu+1\right), \label{Ric:N4}
\end{equation}
gives a Riccati solution of the symmetric form of P$_{\rm IV}$. 
\end{prop}

\subsection{A Determinant formula for a family of classical transcendental solutions}
First, we calculate the Hamiltonians and $\tau$-functions for the
Riccati solution in Proposition \ref{P4:Ric}. 
Under the specialization (\ref{Ric:N4}), the Hamiltonians and
$\tau$-functions are calculated as 
\begin{equation}
h_0=\frac{\varphi_{\nu}'}{\varphi_{\nu}}+\frac{\nu+\frac{1}{2}}{3}t, \quad 
h_1=\frac{\nu+1}{3}t, \quad h_2=\frac{\nu}{3}t, 
\end{equation}
and 
\begin{equation}
\begin{array}{l}
\medskip
\disp 
 \tau_0=\tau_{0,0}=\varphi_{\nu}\exp\left(\frac{\nu+\frac{1}{2}}{6}t^2\right), \quad 
 \tau_1=\tau_{1,0}=\exp\left(\frac{\nu+1}{6}t^2\right), \quad 
 \tau_2=\tau_{1,1}=\exp\left(\frac{\nu}{6}t^2\right), \\
\medskip
\disp 
 s_0(\tau_0)=\tau_{2,1}=0, \\
\disp 
 s_1(\tau_1)=\tau_{0,1}=\varphi_{\nu-1}\exp\left(\frac{\nu-\frac{1}{2}}{6}t^2\right), \quad
 s_2(\tau_2)=\tau_{0,-1}
 =(\nu+1)\varphi_{\nu+1}\exp\left(\frac{\nu+\frac{3}{2}}{6}t^2\right), 
\end{array}
\end{equation}
up to multiplication by some constants, respectively. 
Introducing the functions $\sigma_{k,l}$ by 
\begin{equation}
\tau_{k,l}=\sigma_{k,l}\exp\left(\frac{\nu+2k-l-1}{6}t^2\right), 
\end{equation}
we see that 
\begin{equation}
\sigma_{2,l}=0, \quad 
\sigma_{1,l}=\mbox{const.}, \quad 
\sigma_{0,l}=(\mbox{const.})\times e^{\frac{t^2}{4}}\varphi_{\nu-l}. 
\end{equation}
Moreover, we set 
\begin{equation}
\sigma_{k,l}=\omega_{k,l}\rho_{k,l}, \quad \omega_{k,l}=\omega_{k,l}(\nu), 
\end{equation}
with $\rho_{1,l}=1$ and $\rho_{0,l}=e^{\frac{t^2}{4}}\varphi_{\nu-l}$,
and impose that the constants $\omega_{k,l}$ satisfy 
\begin{equation}
\omega_{k+1,l}\omega_{k-1,l}=\omega_{k,l}^2. \label{omega:k:P4}
\end{equation}
From the Toda equation (\ref{Toda:P4}), the function $\rho_{k,l}$ are
determined by 
\begin{equation}
\rho_{k+1,l}\rho_{k-1,l}=
\rho_{k,l}''\rho_{k,l}-(\rho_{k,l}')^2,
\end{equation}
with the initial conditions 
\begin{equation}
\rho_{2,l}=0, \quad \rho_{1,l}=1, \quad \rho_{0,l}=e^{\frac{t^2}{4}}\varphi_{\nu-l}. 
\end{equation}
By Darboux's formula, the functions $\rho_{1-n,l}$ for $n\in\BZ_{\ge0}$
are expressed as 
\begin{equation}
\rho_{1-n,l}=
 \left|
 \begin{array}{cccc}
   \rho_{0,l}^{(0)}   & \rho_{0,l}^{(1)} & \cdots & \rho_{0,l}^{(n-1)}  \\
   \rho_{0,l}^{(1)}   & \rho_{0,l}^{(2)} & \cdots & \rho_{0,l}^{(n)}    \\
   \vdots             & \vdots           & \ddots & \vdots              \\
   \rho_{0,l}^{(n-1)} & \rho_{0,l}^{(n)} & \cdots & \rho_{0,l}^{(2n-2)}
 \end{array}
 \right|,   \quad \rho_{0,l}^{(i)}=\left(\frac{d}{dt}\right)^i \rho_{0,l}. 
\end{equation}
Note that the constants $\omega_{k,l}$ are determined by the recurrence
relations (\ref{omega:k:P4}) and 
\begin{equation}
\omega_{i,l+1}\omega_{i,l-1}=(\nu+1-l)\omega_{i,l}^2, 
\end{equation}
with the initial conditions $\omega_{i,0}=\omega_{i,1}=1~(i=0,1)$. 
Since it is possible to set $l=0$ without loss of generality, 
we obtain the following theorem. 

\begin{thm}\label{spec:P4}
We define the functions $\tau_n^{\nu-k}$ by (\ref{tau:P4}). 
Then, 
\begin{equation}
\begin{array}{l}
\medskip
\disp
f_0=\frac{\tau_{n+1}^{\nu}\tau_{n-1}^{\nu-1}}
         {\tau_n^{\nu-1}\tau_n^{\nu}}
   =\frac{d}{dt}\log\frac{\tau_n^{\nu-1}}{\tau_n^{\nu}}, \\
\medskip
\disp
f_1=\frac{\tau_n^{\nu}\tau_{n+1}^{\nu-1}}
         {\tau_{n+1}^{\nu}\tau_n^{\nu-1}}
   =\frac{d}{dt}\log\frac{\tau_{n+1}^{\nu}}{\tau_n^{\nu-1}}, \\
\disp
f_2=(\nu+1)
    \frac{\tau_n^{\nu-1}\tau_{n+1}^{\nu+1}}
         {\tau_n^{\nu}\tau_{n+1}^{\nu}}
   =\frac{d}{dt}\log\frac{\tau_n^{\nu}}{\tau_{n+1}^{\nu}}+t, 
\end{array}
\end{equation}
\begin{equation}
(\aa_0,\aa_1,\aa_2)=\left(-n,-\nu+n,\nu+1\right), 
\end{equation}
and 
\begin{equation}
q=-\frac{\tau_n^{\nu}\tau_{n+1}^{\nu-1}}
        {\tau_{n+1}^{\nu}\tau_n^{\nu-1}}
 =-\frac{d}{dt}\log\frac{\tau_{n+1}^{\nu}}{\tau_n^{\nu-1}}, \quad 
p=(\nu+1)
  \frac{\tau_n^{\nu-1}\tau_{n+1}^{\nu+1}}
       {\tau_n^{\nu}\tau_{n+1}^{\nu}}
 =\frac{d}{dt}\log\frac{\tau_n^{\nu}}{\tau_{n+1}^{\nu}}+t, 
\end{equation}
\begin{equation}
\ka_0=-\nu+n, \quad \theta_{\infty}=-\nu-1, 
\end{equation}
gives a family of classical transcendental solutions of the symmetric
form of P$_{\rm IV}$ and the Hamiltonian system S$_{\rm IV}$,
respectively. 
\end{thm}

\end{document}